\documentclass[12pt]{article}
\usepackage[margin=1in]{geometry}
\usepackage[round]{natbib}
\usepackage{amsmath,amssymb,authblk,bm,enumerate,graphicx,rotating}

\def\bbeta{\boldsymbol{\beta}}
\def\expit{\mathrm{expit}}
\def\logit{\mathrm{logit}}
\def\btheta{\boldsymbol{\theta}}
\def\pr{\mathrm{P}}
\def\T{\mathbf{T}}

\def\X{\mathbf{X}}

\DeclareMathOperator*{\argmax}{arg\,max}
\DeclareMathOperator*{\argmin}{arg\,min}

\title{Integrative genetic risk prediction using nonparametric empirical Bayes classification}
\author[1]{Sihai Dave Zhao}
\affil[1]{Department of Statistics, University of Illinois at Urbana-Champaign}

\begin{document}
\maketitle

\begin{abstract}
Genetic risk prediction is an important component of individualized medicine, but prediction accuracies remain low for many complex diseases. A fundamental limitation is the sample sizes of the studies on which the prediction algorithms are trained. One way to increase the effective sample size is to integrate information from previously existing studies. However, it can be difficult to find existing data that examine the target disease of interest, especially if that disease is rare or poorly studied. Furthermore, individual-level genotype data from these auxiliary studies are typically difficult to obtain. This paper proposes a new approach to integrative genetic risk prediction of complex diseases with binary phenotypes. It accommodates possible heterogeneity in the genetic etiologies of the target and auxiliary diseases using a tuning parameter-free nonparametric empirical Bayes procedure, and can be trained using only auxiliary summary statistics. Simulation studies show that the proposed method can provide superior predictive accuracy relative to non-integrative as well as integrative classifiers. The method is applied to a recent study of pediatric autoimmune diseases, where it substantially reduces prediction error for certain target/auxiliary disease combinations. The proposed method is implemented in the R package \verb|ssa|.
\end{abstract}

\section{Introduction}
Genetic risk prediction for complex diseases is an important but difficult problem. Genome-wide association studies (GWAS) have successfully identified many SNPs associated with human disease, but it has been difficult to translate these successes into accurate risk prediction models \citep{kraft2009genetic,jostins2011genetic}. The low accuracies can in part be attributed to the fact that much of the heritability of a complex disease is likely due to a large number of SNPs whose effect sizes are too weak to be discovered in any given GWAS \citep{manolio2009finding,chatterjee2013projecting,dudbridge2013power}.

To address this issue, recent research has focused on developing new prediction algorithms that aggregate information over a large number of SNPs, rather than using only those that reach genome-wide significance \citep{chatterjee2016developing}. For example, polygenic risk scores can be constructed by taking weighted sums of all typed SNPs, or all SNPs that pass a loose significance threshold. The weights can be calculated based on univariate regression coefficients \citep{purcell2009common,chatterjee2013projecting,shi2016winners}, with or without accounting for linkage disequilibrium \citep{vilhjalmsson2015modeling,mak2016polygenic}, or by using machine learning algorithms such as the lasso \citep{wei2013large,okser2014regularized}. The weights can also be treated as random draws from some prior distribution, and risk scores can be constructed using their posterior distributions \citep{zhou2013polygenic,golan2014effective,speed2014multiblup}.

These new methods are still fundamentally limited by the sample sizes of the GWAS data on which they are trained \citep{wray2013pitfalls}. A straightforward way to increase sample size is to recruit additional study subjects, but this is time-consuming and costly. Instead, the training data can be augmented with data from previously existing studies of the disease of interest, for example by using meta-analytic methods \citep{stahl2012bayesian,cross2013genetic,cross2013identification,shi2016novel}. However, this is only feasible when developing risk prediction models for well-studied diseases. For other diseases, especially rare conditions, there may not be many existing GWAS studies, and the ones that do exist may be so small that integrating them may not be very useful.

This paper studies an alternative method of increasing effective sample size: borrowing information from auxiliary GWAS studies of diseases different from, but potentially related to, the target disease of interest. For example, recent studies of co-heritability have uncovered high degrees of genetic correlation between psychiatric disorders \citep{lee2012estimating,yang2013polygenic} and between autoimmune diseases \citep{li2015meta,li2015genetic}. Genetic correlation between the target and auxiliary diseases implies that some of the SNPs that are predictive of one will simultaneously be predictive of others. This dramatically enlarges the pool of existing studies that can be leveraged to improve prediction accuracy for the target disease. A major difficulty is that individual-level genotype data from existing studies are often difficult to obtain due to privacy concerns, so prediction methods that can be trained using only GWAS summary statistics are preferable.

So far, it appears that there exist very few methods capable of this type of integrative genetic risk prediction. Given GWAS data for a target disease and auxiliary diseases, \citet{li2014improving} and \citet{maier2015joint} posit linear models for the effects of the SNPs, so that to each SNP there corresponds a vector of regression coefficients, one for each disease. They then assume a multivariate prior distribution on these coefficients; the degree of correlation between the coefficients quantifies the amount of information that can be borrowed across diseases. However, they require parametric assumptions on the prior distribution as well as selection or estimation of tuning parameters, e.g. those that govern the prior covariances between the coefficients, which can be computationally intensive and inaccurate. Furthermore, their methods require raw genotype data. There do not appear to exist any genetic risk prediction methods than can integrate only summary statistics from auxiliary GWAS.

This paper proposes a new approach to integrative genetic risk prediction of complex diseases with binary phenotypes. It does not require raw genotype data from either the disease of interest or the related diseases and can be trained using only summary statistics. It also uses a tuning parameter-free nonparametric empirical Bayes procedure to estimate prior distributions; this automatically learns the degree of genetic similarity between the target and auxiliary diseases. The proposed method is computationally straightforward and is implemented in the R package \verb|ssa|.

The remainder of this paper is organized as follows. Section~\ref{sec:methods} describes the proposed method and Section~\ref{sec:results} studies its performance in simulations and in a study of pediatric autoimmune diseases, conducted by Hakonarson and colleagues at the Children's Hospital of Philadelphia \citep{li2015meta,li2015genetic}. Section~\ref{sec:discussion} concludes with a discussion of possible extensions and future work.

\section{\label{sec:methods}Methods}
\subsection{\label{sec:formulation}Statistical formulation}
Genetic risk prediction for binary disease phenotypes can be formulated as a classification problem. For a new subject whose disease status $Y_{new}\in\{0,1\}$ is unobserved, let $\X_{new}=(X_{new,1},\ldots,X_{new,d})^\top$ be the observed genotypes of $d$ SNPs, where $X_{new,j}$ is the number of minor alleles of the $j$th SNP. The goal is to use $\X_{new}$ to predict $Y_{new}$, where for example $Y_{new}=0$ means that the subject does not have the disease.

Let $\mathcal{D}=\{(\X_i,Y_i),i=1,\ldots,n\}$ denote the training data for the target disease, where $\X_i=(X_{i1},\ldots,X_{id})^\top$ is the vector of genotypes of the $i$th subject, $n_0$ and $n_1$ are the numbers of training subjects with $Y_i=0$ and $Y_i=1$, respectively, and $n=n_0+n_1$. Also assume that summary statistics from an auxiliary GWAS study of another disease, potentially related to the one of interest, are available. Denote these statistics by $\T=(T_1,\ldots,T_d)^\top$, where $T_j$ is the test statistic for the marginal association between the $j$th SNP and the auxiliary disease. For clarity, it will be assumed that only a single auxiliary GWAS is used. It is conceptually straightforward to extend the proposed method to multiple auxiliary GWAS results; this is further discussed in Section~\ref{sec:discussion}.

The integrative genetic risk prediction problem is to develop a classifier
\[
\delta(\X_{new};\mathcal{D},\T):\mathbb{R}^d\times(\mathbb{R}^d\times\{0,1\})^n\times\mathbb{R}^d\rightarrow\{0,1\},
\]
trained using both $\mathcal{D}$ and $\T$, that minimizes the misclassification rate
\begin{equation}
  \label{eq:R}
  R(\delta)=\pr\{Y_{new}\ne\delta(\X_{new};\mathcal{D},\T)\}.
\end{equation}
This differs from the standard non-integrative risk prediction problem, where the classifier $\delta$ is allowed to depend only on $\mathcal{D}$ and not on $\T$.

Several assumptions are made throughout the remainder of this paper. First, the $(\X_i,Y_i)$ are assumed to be independent and identically distributed across all $i=new,1,\ldots,n$. Next, all SNPs in $\X_i$ are assumed to be in linkage equilibrium. This can be approximately achieved by using linkage disequilibrium (LD) pruning. There is evidence that pruning can improve the accuracy of genetic risk prediction \citep{shi2016winners}, and even if pruning is not done, ignoring LD and treating SNPs as independent can still give accurate classification \citep{bickel2004some,hand2006classifier,zhao2014menos}. A complete treatment of classification under LD is difficult and is left for future work.

Finally, all SNPs are assumed to be in Hardy-Weinberg equilibrium, so that for $i=new,1,\ldots,n$,
\[
X_{ij}\mid Y_i=y\sim Bin(2,\pi_{yj}),\quad y\in\{0,1\},
\]
where $\pi_{yj}$ is the minor allele frequency of the $j$th SNP in class $y$. Typed SNPs that are not in Hardy-Weinberg equilibrium can be dropped from analysis. The auxiliary summary statistics $T_j$ are assumed to arise from chi-square statistics from an existing GWAS study conducted on an independent sample of subjects, so the $T_j$ are statistically independent of $(X_{new},Y_{new})$ and $\mathcal{D}$ and follow
\[
T_j\sim\chi^2_1(\lambda_j).
\]
SNPs that are not associated with the auxiliary disease will have $\lambda_j=0$. Many common association tests, such as the allelic test for association, give chi-square test statistics, and the approach proposed below can be easily modified if the $T_j$ are otherwise distributed.

\subsection{\label{sec:classification}Review of non-integrative classification}
It is well-known \citep{devroye1996probabilistic} that the optimal classifier that minimizes the misclassification rate $R(\delta)$~\eqref{eq:R} is given by
\[
\delta^\star(\X_{new};\mathcal{D},\T)
=
I\left\{
\frac{\pr(Y_{new}=1\mid\X_{new},\mathcal{D},\T)}{\pr(Y_{new}=0\mid\X_{new},\mathcal{D},\T)}
\geq
1
\right\}.
\]
Under the assumptions in Section~\ref{sec:formulation}, $(X_{new},Y_{new})$ is independent of $\mathcal{D}$ and $\T$, so the optimal classifier reduces to
\begin{equation}
  \label{eq:oracle}
  \delta^\star(\X_{new};\mathcal{D},\T)
  =
  I\left\{
  \log\frac{P}{1-P}
  +
  \sum_{j=1}^d
  \log\frac{f_{Bin}(X_{new,j};2,\pi_{1j})}{f_{Bin}(X_{new,j};2,\pi_{0j})}
  \geq
  0
  \right\},
\end{equation}
where $P=\pr(Y_{new}=1)$ is the prevalence of the target disease of interest and $f_{Bin}(x;2,\pi)$ is the probability mass function of a $Bin(2,\pi)$ random variable.

The form of $\delta^\star$~\eqref{eq:oracle} shows that optimal prediction of the target disease does not benefit from integration of the $\T$. This is one reason why standard methods for genetic risk prediction do not consider auxiliary sources of information. Of course, optimal prediction also does not benefit from the training data $\mathcal{D}$ either, because the optimal classifier uses the true minor allele frequencies $\pi_{yj}$. It makes sense that if the true $\pi_{yj}$ were known, both training data and auxiliary GWAS summary statistics would be irrelevant for optimal prediction.

Clearly, the oracle $\delta^\star$ cannot be implemented in practice, since the $\pi_{yj}$ are unknown. Instead, most existing classifiers calculate estimates of $\pi_{yj}$ using $\mathcal{D}$, which are then plugged into $\delta^\star$. Using maximum likelihood estimates of $\pi_{yj}$ leads to the standard naive Bayes classifier. However, since the total number of SNPs $d$ is large, maximum likelihood estimation of the high-dimensional vectors $(\pi_{y1},\ldots,\pi_{yd})^\top$ for $y\in\{0,1\}$ can be inaccurate. A popular alternative is to use some form of regularized estimation. A number of strategies for high-dimensional discriminant analysis have been developed in this vein, though mostly under the assumption that the $X_{ij}$ are normal rather than binomial random variables \citep{fan2008high,greenshtein2009application,cai2012direct,fan2012road,mai2012direct,fan2013optimal,han2013coda,dicker2016nonparametric}.

\subsection{\label{sec:bayes_model}Integrative classification via Bayesian modeling}
It is not obvious how to properly incorporate the auxiliary summary statistics $T_j$ into an integrative classifier. The form of the optimal classifier $\delta^\star$~\eqref{eq:oracle} gives no indication as to how the $T_j$ can be used, yet it is intuitively clear that the parameters $\lambda_j$ underlying the $T_j$ can contain information about the $\pi_{yj}$ and should be leveraged. The $\lambda_j$ can be viewed as latent annotation information for each SNP. If the target disease is truly related to the auxiliary disease, a large value of $\lambda_j$ provides additional evidence that the $j$th SNP may be useful for predicting $Y_{new}$, even if the effect size of that SNP in the target disease training data $\mathcal{D}$ is weak.

Properly leveraging the $T_j$ to improve prediction of the target disease poses several methodological challenges. First, the $\lambda_j$ are not directly observed. Second, how they should be used depends on the extent of the genetic similarity between the target and auxiliary diseases. For example, some SNPs may be predictive only of the auxiliary disease but not of the target disease, or vice versa, so just because a SNP has a large $\lambda_j$ in the auxiliary GWAS does not necessarily mean that it is useful for predicting the target disease. It is not clear how best to leverage the $\lambda_j$ in this case. Finally, it may not always be known whether genetic correlations exist, for example if the diseases are poorly understood.

This paper proposes a new method that can address each of these challenges. The method is motivated by a Bayesian model for the $(X_{new},Y_{new})$, $\mathcal{D}$, and $\T$, which assumes that
\begin{equation}
  \label{eq:G}
  (\pi_{0j},\pi_{1j},\lambda_j)\sim G
\end{equation}
for some trivariate prior distribution $G$. Under this assumption, $(X_{new},Y_{new})$, $\mathcal{D}$, and $\T$ are no longer necessarily mutually independent, and the misclassification rate $R(\delta)$~\eqref{eq:R} is be minimized by the classifier
\begin{equation}
  \label{eq:int}
  \begin{aligned}
    &I\left\{
    -\log\frac{P}{1-P}\,
    \leq\right.\\
    &\left.
    \sum_{j=1}^d\log
    \frac{\int f_{Bin}(X_{new,j};2,u_1)f_{Bin}(S_{0j};2n_0,u_0)f_{Bin}(S_{1j};2n_1,u_1)f_{\chi^2}(T_j;1,l)dG(u_0,u_1,l)}
         {\int f_{Bin}(X_{new,j};2,u_0)f_{Bin}(S_{0j};2n_0,u_0)f_{Bin}(S_{1j};2n_1,u_1)f_{\chi^2}(T_j;1,l)dG(u_0,u_1,l)}
     \right\},
  \end{aligned}
\end{equation}
where $S_{yj}=\sum_{i:Y_i=y}X_{ij}$ and $f_{\chi^2}(x;\nu,\lambda)$ is the probability density function of a $\chi^2_\nu(\lambda)$ distribution. To derive~\eqref{eq:int} it is also assumed that the $(\pi_{0j},\pi_{1j},\lambda_j)$ and the $Y_i,i=new,1,\ldots,n$ are independent, which is reasonable because the parameter values can be thought of as being drawn from $G$ independently of the cases and controls being drawn from the population of subjects.

The form of~\eqref{eq:int} is the key to the proposed approach. Unlike the optimal classifier $\delta^\star$~\eqref{eq:oracle}, in which neither $\T$ nor $\mathcal{D}$ appear,~\eqref{eq:int} provides a sensible way for integrating the $T_j$ with the $\mathcal{D}$. In fact, if the Bayesian assumption~\eqref{eq:G} is true,~\eqref{eq:int} is the optimal method of integrating $T_j$ and $\mathcal{D}$. Even under the present frequentist setting described in Section~\ref{sec:formulation}, procedures motivated by Bayesian formalisms can still have excellent performance, for example in frequentist compound decision problems \citep{robbins1951asymptotically,robbins1956empirical,zhang2003compound,brown2009nonparametric,jiang2009general,gu2015problem}.

The prior $G$ implicitly encodes the additional information about $(\pi_{0j},\pi_{1j})$ that can be borrowed from the auxiliary study, via the correlation between the minor allele frequencies and the latent annotations $\lambda_j$. In the extreme case where the target and auxiliary diseases are not genetically correlated, $G$ factors into the product of a bivariate distribution on $(\pi_{0j},\pi_{1j})$ and a univariate distribution on $\lambda_j$. The terms involving $T_j$ will then cancel out in~\eqref{eq:int}, resulting in a non-integrative classifier that depends only on the target training data.

\subsection{\label{sec:neb}Nonparametric empirical Bayes classification using latent annotations}
The integrals in~\eqref{eq:int} must be estimated, because the prior $G$ is unknown. One possibility is to assume that $G$ lies in a parametric family. However, $G$ is a complex multivariate distribution and it is not clear what family can be used. An attractive alternative is the nonparametric maximum likelihood estimator of \citet{kiefer1956consistency}, which in the present context takes the form
\begin{equation}
  \label{eq:npmle}
  \hat{G}
  =
  \argmax_{G\in\mathcal{G}}
  \prod_{j=1}^d\int
  f_{Bin}(S_{0j};2n_0,u_0)f_{Bin}(S_{1j};2n_1,u_1)f_{\chi^2}(T_j;1,l)dG(u_0,u_1,l),
\end{equation}
where $\mathcal{G}$ is the set of all trivariate distributions. The advantage of $\hat{G}$ is that it requires minimal assumptions, no tuning parameters, and is a consistent estimator of the true mixing distribution $G$ \citep{kiefer1956consistency}. More importantly, it uses the observed data to automatically learn the degree to which the target and auxiliary diseases are related.

The proposed classifier is obtained by replacing $G$ in the Bayesian classifier~\eqref{eq:int} with the estimate $\hat{G}$. This will be referred to as a nonparametric empirical Bayes classifier using latent annotations, or NEBULA. It is clear from~\eqref{eq:int} and~\eqref{eq:npmle} that NEBULA can be trained using only summary statistics $T_j$ from the auxiliary GWAS and $S_{yj}$ from the target disease training data. This is a major advantage over existing integrative classifiers. NEBULA can also incorporate additional non-genetic predictors such as age or gender. Denote these by $Z_{new,j}$, $j=1,\ldots,q$ and let $f_j(z;\btheta_{yj})$ be the density of the $j$th predictor in class $y\in\{0,1\}$, where $\btheta_{yj}$ is a vector of parameters. Let $\hat{\btheta}_{yj}$ be the maximum likelihood estimates of $\btheta_{yj}$ obtained from training data $Z_{ij},i=1,\ldots,n$. With these additional covariates, NEBULA is defined as
\begin{equation}
  \label{eq:nebula}
  \begin{aligned}
    &\hat{\delta}_{NEBULA}(X_{new};\mathcal{D},\T)
    =
    I\left\{
    -\log\frac{P}{1-P}
    \leq
    \sum_{j=1}^q\log\frac{f_j(Z_{new,j};\hat{\btheta}_1)}{f_j(Z_{new,j};\hat{\btheta}_0)}\,+
    \right.\\
    &\left.
    \sum_{j=1}^d\log
    \frac{\int f_{Bin}(X_{new,j};2,u_1)f_{Bin}(S_{0j};2n_0,u_0)f_{Bin}(S_{1j};2n_1,u_1)f_{\chi^2}(T_j;1,l)d\hat{G}(u_0,u_1,l)}
         {\int f_{Bin}(X_{new,j};2,u_0)f_{Bin}(S_{0j};2n_0,u_0)f_{Bin}(S_{1j};2n_1,u_1)f_{\chi^2}(T_j;1,l)d\hat{G}(u_0,u_1,l)}
     \right\}.
  \end{aligned}
\end{equation}
NEBULA is designed to directly predict the class $Y_{new}$ of the newly observed genotype vector $\X_{new}$, but the sum inside the indicator function in~\eqref{eq:nebula} can also be treated as a continuous score for the newly observed subject. This score can then be used to calculate area under the curve statistics or can be calibrated to provide a risk score \citep{cook2007use}.

The accuracy of NEBULA is determined in part by the sample sizes of the target and auxiliary studies. Larger sample sizes reduce the signal-to-noise ratios of the $S_{0j}$, $S_{1j}$, and $T_j$. For example, for $y\in\{0,1\}$, the inverse coefficient of variation of the binomial random variable $S_{yj}$ is $(n_y\pi_{yj}^{1/2})/\{n_y\pi_{yj}(1-\pi_{yj})\}^{-1/2}$, which increases with the square root of the target study sample size. Similarly, the inverse of the coefficient of variation of the noncentral chi-square $T_j$ is $(1+\lambda_j)/\{2(k+2\lambda_j)\}^{1/2}$. Since SNPs that are associated with the auxiliary disease have noncentrality parameters $\lambda_j$ that are directly proportional to the sample size of the auxiliary study, this quantity increases with the square root of the auxiliary sample size. These smaller signal-to-noise ratios make the deconvolution problem less difficult and thus allow for better estimate of the integrals in~\eqref{eq:int}.


\subsection{\label{sec:imp}Implementation}
The Kiefer-Wolfowitz estimator $\hat{G}$~\eqref{eq:npmle} is difficult to calculate. To mitigate the computational burden, \citet{koenker2014convex} recently proposed a finite-dimensional approximation to $\hat{G}$. For $y\in\{0,1\}$ let $\Pi_y$ be a set of $d_y$ equally-spaced grid points $\min_j\hat{\pi}_{yj}=u_{y1}<\ldots<u_{yd_y}=\max_j\hat{\pi}_{yj}$, where $\hat{\pi}_{yj}=S_{yj}/(2n_y)$ is the maximum likelihood estimate of $\pi_{yj}$. Similarly, let $\Lambda$ be a set of $d_2$ equally-spaced grid points $\min_jT_j=l_1<\ldots<l_{d_2}=\max_jT_j$. The Koenker-Mizera approximation is identical to~\eqref{eq:npmle} except that the optimization is not over the class of all trivariate distributions, but instead over all discrete trivariate distributions supported on $\Pi_0\times\Pi_1\times\Lambda$. The resulting estimator is the solution to a discrete convex optimization problem and can be conveniently computed \citep{feng2016nonparametric}.
  
Though the Koenker-Mizera estimator is an approximation to the true nonparametric maximum likelihood estimator $\hat{G}$, it has been shown to work extremely well in many problem \citep{koenker2014frailty,koenker2014gaussian,gu2015unobserved,dicker2016nonparametric,feng2016nonparametric,jiang2016generalized}. Ideally the number of points $d_0$, $d_1$, and $d_2$ used to construct the grids $\Pi_0$, $\Pi_1$, and $\Lambda$ should be as large as possible given constraints on computation time and memory, but in practice relatively few are needed for good performance. A theoretical characterization of a sufficient number of grid points is given by \citet{dicker2016nonparametric}.

Koenker-Mizera estimators for various problems are calculated using fast interior point methods in the R package \verb|REBayes| \citep{koenker2013rebayes,koenker2014convexR}. However, for the trivariate problem~\eqref{eq:npmle} considered in this paper, \verb|REBayes| does not contain a ready-made implementation. The expectation-maximization algorithm offers a simple alternative \citep{feng2016nonparametric}. Specifically, let $dG^k(u_0,u_1,l)$ be the mass corresponding to the grid point $(u_0,u_1,l)\in\Pi_0\times\Pi_1\times\Lambda$ at the $k$th iteration of the algorithm. Then the $k+1$th update is
\begin{equation}
  \label{eq:km}
  \begin{aligned}
    &dG^{k+1}(u_0,u_1,l)
    =\\
    &\frac{1}{d}\sum_{j=1}^d
    \frac{dG^{k}(u_0,u_1,l)f_{Bin}(S_{0j};2n_0,u_0)f_{Bin}(S_{1j};2n_1,u_1)f_{\chi^2}(T_j;1,l)}
         {\sum_{(u_0,u_1,l)\in\Pi_0\times\Pi_1\times\Lambda}dG^{k}(u_0,u_1,l)f_{Bin}(S_{0j};2n_0,u_0)f_{Bin}(S_{1j};2n_1,u_1)f_{\chi^2}(T_j;1,l)}.
  \end{aligned}
\end{equation}
This implementation has been made available in the R package \verb|ssa|.

\section{\label{sec:results}Results}
\subsection{\label{sec:compared}Methods compared}
In order to provide a performance baseline against which NEBULA~\eqref{eq:nebula} can be judged, the standard non-integrative polygenic risk score (PRS) classifier \citep{purcell2009common,shi2016novel} was implemented. Let $P$ be the prevalence of the target disease in the population. In practice $P$ is often known from previous epidemiological studies, or in cohort sampling designs can be estimated from the training data. The PRS classifier is then defined as
\begin{equation}
  \label{eq:prs}
  \delta_{PRS}(\X_{new};\mathcal{D})
  =
  I\left\{
  2\sum_{j=1}^d\log\frac{1-\hat{\pi}_{1j}}{1-\hat{\pi}_{0j}}
  +
  \sum_{j=1}^d\hat{\beta}_jI(\vert\hat{\beta}_j\vert>\lambda)X_{new,j}
  \geq
  -\log\frac{P}{1-P}
  \right\},
\end{equation}
where
\[
\hat{\beta}_j=\log\frac{\hat{\pi}_{1j}/(1-\hat{\pi}_{1j})}{\hat{\pi}_{0j}/(1-\hat{\pi}_{0j})}
\]
is the maximum likelihood estimate of the log-odds ratio based on the training data $\mathcal{D}$. The parameter $\lambda$ is a threshold that serves to remove unimportant SNPs from the classifier. The PRS is nearly identical to the oracle classifier $\delta^\star$~\eqref{eq:oracle} after plugging in the maximum likelihood estimates of the $\pi_{yj}$, except for the thresholding step. The optimal value of $\lambda$ can be determined using cross-validation.

In addition to NEBULA, two new alternative integrative genetic risk prediction algorithms were implemented for comparison. Neither of these has been yet been described in the literature and may be interesting in their own rights. Let $\hat{\gamma}_j$ denote the log-odds ratio of the $j$th SNP estimated from the auxiliary GWAS data. SNPs with larger $T_j$ will also have $\hat{\gamma}_j$ with larger magnitudes. The new classifiers are:
\begin{enumerate}
\item An adaptive version of the PRS: SNPs with $\vert\hat{\beta}_j\hat{\gamma}_j\vert\leq\lambda$ are dropped; the remaining ones are still weighted by $\hat{\beta}_j$ in the PRS. This allows SNPs with relatively weak $\vert\hat{\beta}_j\vert$ to still be included in the prediction algorithm as long as $\vert\hat{\gamma}_j\vert$ is sufficiently large.

\item Adaptive lasso \citep{zou2006adaptive}: logistic lasso regression \citep{tibshirani1996regression} is implemented using the $\hat{\gamma}_j$ as adaptive weights:
  \[
  \hat{\bbeta}=\argmin_{\bbeta}\left\{-\ell(\bbeta)+\lambda\sum_{j=}^d\frac{\vert\beta_j\vert}{\vert\hat{\gamma}_j\vert}\right\},
  \]
  where $\bbeta=(\beta_1,\ldots,\beta_d)^\top$ is the regression coefficient, $\ell(\bbeta)$ is logistic log-likelihood of the training data $\mathcal{D}$, and $\lambda$ is a tuning parameter selected using cross-validation. This has the same effect as the adaptive PRS.
\end{enumerate}
A potential problem with these alternative methods is that they implicitly assume that the SNPs that are significant in the auxiliary data are the same as those that are predictive of the target disease. When this is indeed true, or approximately true, these methods should perform well, but heterogeneity between the genetic etiologies of the target and auxiliary diseases may cause these methods to have low prediction accuracy.

\subsection{\label{sec:sims}Simulation studies}
To simulate genotype data for the target disease of interest, the total number of SNPs was $d=10,000$ in all simulations. Some SNPs were set to be associated with the target disease; the number of these non-null SNPs was varied across simulation settings. In controls, the minor allele frequencies $\pi_{0j}$ were randomly generated uniformly in $[0.2,0.5]$. In cases, the minor allele frequencies for null SNPs were set equal to the corresponding $\pi_{0j}$. For non-null SNPs, log-odds ratios $\beta_j$ were first generated from $N(\mu,0.01)$ and then randomly multiplied by 1 or $-1$. The value of $\mu$ was varied across simulation settings. The minor allele frequencies were then set to $\pi_{1j}=\expit\{\beta_j+\logit(\pi_{0j})\}$, where $\expit(x)$ is the inverse of the $\logit(x)=\log\{x/(1-x)\}$. Once generated, the $\pi_{0j}$ and $\pi_{1j}$ were fixed across all replications. Finally, to generate the training data $\mathcal{D}$, $n_0=n_1=100$ control and case genotype vectors were generated with $X_{ij}\sim Bin(2,\pi_{0j})$ and $X_{ij}\sim Bin(2,\pi_{1j})$, respectively. To generate the testing data, $n_0=n_1=50$ controls and cases were generated in the same way. A prevalence of $P=1/2$ was used when implementing all classifiers.

Genotype data for the auxiliary disease were simulated in the same fashion, but with different sample sizes. The number of SNPs associated with the auxiliary disease was also varied across simulation settings. The percent overlap between the SNPs that were non-null for the auxiliary disease and those that were non-null for the target disease was varied as well, representing different degrees of genetic heterogeneity between the diseases. When the two sets of non-null SNPs had different sizes, the percentage is understood to be relative to the smaller set. Estimated log-odds ratios $\hat{\gamma}_j$ and chi-square test statistics $T_j$ from allelic tests of association were calculated from the auxiliary data. Only these summary statistics, and not the raw data, were used by the integrative algorithms described in Section~\ref{sec:compared}. The Koenker-Mizera estimator~\eqref{eq:km} used in the NEBULA classifier was implemented using 20 equally-spaced points to construct the grids $\Pi_y,y\in\{0,1\}$ and $\Lambda$.

\begin{figure}
  \centerline{\includegraphics[width=\textwidth]{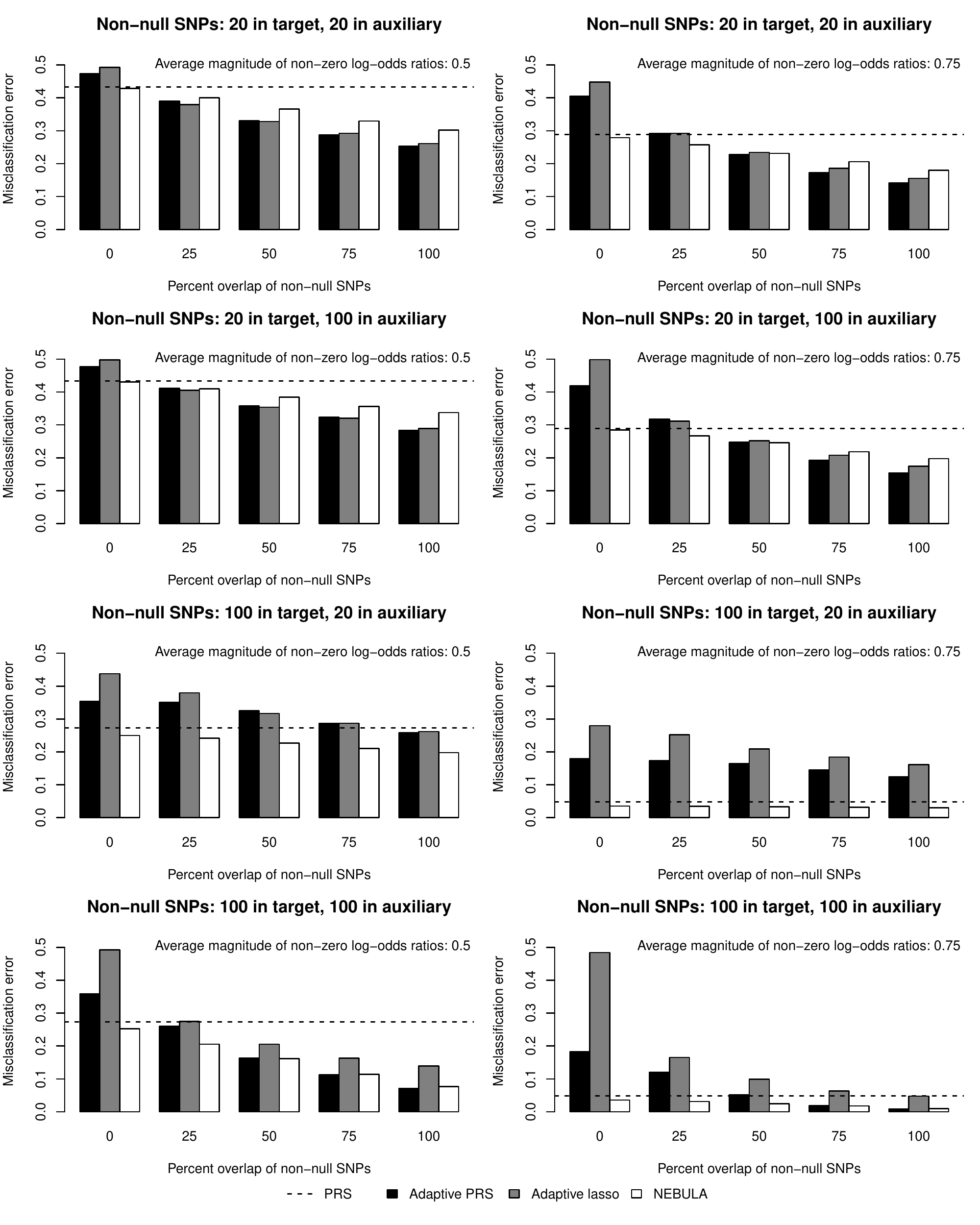}}
  \caption{\label{fig:sims}Average misclassification rates when the auxiliary data were generated with $n_0=n_1=1,000$ controls and cases.}
\end{figure}

Figure~\ref{fig:sims} reports the average misclassification rates across 200 replications of each simulation setting when the auxiliary data were generated with $n_0=n_1=1,000$ controls and cases. All three integrative classifiers had improved accuracy as the percentage overlap of non-null SNPs increased. This confirms that each method was indeed able to utilize information from the auxiliary summary statistics.

\begin{figure}
  \centering
  \includegraphics[height=0.9\textheight]{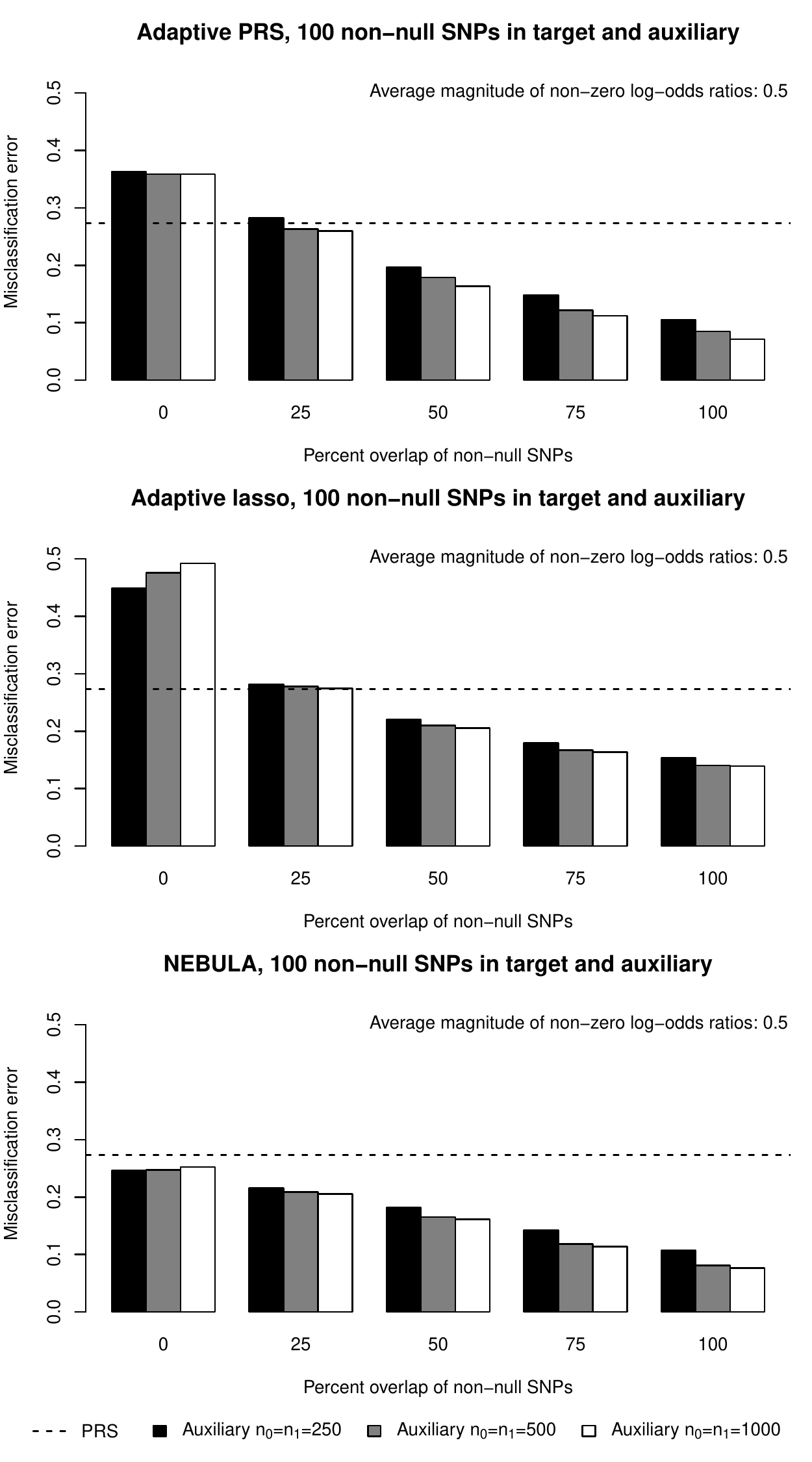}
  \caption{\label{fig:samplesize}Average misclassification rates for different auxiliary study sample sizes}
\end{figure}

Figure~\ref{fig:samplesize} illustrates the impact of generating 250, 500, and 1,000 cases and controls in the auxiliary data. As discussed in Section~\ref{sec:neb}, larger sample sizes allow for more accurate estimation of the integrals in~\eqref{eq:int}. Not surprisingly, misclassification errors decreased as the size of the auxiliary study increased for all three integrative classifiers.

NEBULA outperformed the other integrative classifiers when the target and auxiliary diseases were more genetically heterogeneous. When the overlap between the target-significant SNPs and the auxiliary-significant SNPs was only 25\%, NEBULA had the lowest misclassification rates in most of the simulation settings. As discussed in Section~\ref{sec:compared}, the adaptive PRS and lasso methods should have difficulty in this situation. In contrast, NEBULA automatically learns the true degree of similarity between the target disease and the auxiliary summary statistics, via the nonparametric maximum likelihood estimator~\eqref{eq:npmle}, and can ignore the auxiliary data if the similarity is not strong. Furthermore, Figure~\ref{fig:sims} shows that NEBULA can outperform the baseline non-integrative PRS even when the target and auxiliary diseases are quite genetically different, while the misclassification rates of the other integrative classifiers could be much higher than baseline. This is important because in practice the degree of similarity between the target and auxiliary diseases may be unknown, especially for poorly understood diseases.

These simulation results also indicate that the relative improvement of NEBULA over the other integrative classifiers was highest when there were more predictive SNPs in the target study. This is because the soft thresholding in the adaptive PRS and adaptive lasso algorithms make them better suited to sparse settings, where there are a small number of important predictors with relatively strong effect sizes. In contrast, NEBULA does not threshold any predictors and instead aggregates information from every SNP. It is thus more suited for dense settings, where there are a large number of important predictors with relatively weak effect sizes. This type of dense configuration may be especially relevant in genetic risk prediction, as it has been hypothesized to be one of the causes of missing heritability \citep{manolio2009finding,chatterjee2013projecting,dudbridge2013power}.

\subsection{Application to pediatric autoimmune disease}
Hakonarson and colleagues at the Children's Hospital of Pennsylvania \citep{li2015meta,li2015genetic} collected genotype data for ten different pediatric autoimmune disorders from more than 5,000 cases and 10,000 shared age- and gender-matched controls of European ancestry; see Table~\ref{tab:paid}. The numbers of cases per disorder were relatively small, ranging from 100 to 2,000, so training genetic risk prediction models is difficult. In this section, the proposed NEBULA classifier~\eqref{eq:nebula} and the alternative methods described in Section~\ref{sec:compared} are applied to this pediatric autoimmune data.

\begin{table}
  \centering
  \caption{\label{tab:paid}Pediatric autoimmune disorders studied by Hakonarson and colleagues.}
  \begin{tabular}{rl}
    \hline
    Disorder & Abbreviation \\
    \hline
    Ankylosing spondylitis & AS \\
    Common variable immunodeficiency disorder & CVID \\
    Crohn's disease & CD \\
    Celiac's disease & CEL \\
    Juvenile idiopathic arthritis & JIA \\
    Psoriasis & PS \\
    Systemic lupus erythematosus & SLE \\
    Thyroiditis & THY \\
    Type I diabetes & T1D \\
    Ulcerative colitis & UC \\
    \hline
  \end{tabular}
\end{table}

To apply the integrative classifiers, each of the ten autoimmune disorders was treated in turn as the target disease for genetic risk prediction, and each of the remaining nine disorders was in turn used as the auxiliary disease. Some pediatric autoimmune diseases have been found to be genetically correlated \citep{li2015meta,li2015genetic}, so it is reasonable to integrate results from one disorder to help predict another. For each target disease, training data consisted of 90\% of the available target disease samples and the same number of subjects from the control data, all randomly selected. The other 10\% of the target disease samples and an equal number of random control subjects comprised the testing data. The remaining control subjects were combined with the data for the auxiliary disease in order to calculate log-odds ratios $\hat{\gamma}_j$ and summary statistics $T_j$. This sample splitting procedure was repeated 50 times and the misclassification errors of the different classifiers were averaged across the replications.

\begin{table}
  \centering
  \caption{\label{tab:egg}Auxiliary GWAS from the EGG Consortium.}
  \begin{tabular}{rccl}
    \hline
    Trait & Abbreviation & Sample size & Reference \\
    \hline
    Birth length & BL & 28,459 & \citet{van2015novel} \\
    Birth weight & BW & 26,836 & \citet{horikoshi2013new}\\
    Childhood BMI & BMI & 35,668 & \citet{felix2016genome}\\
    Childhood obesity & OB & 13,848 & \citet{bradfield2012genome} \\
    \hline
  \end{tabular}
\end{table}

In addition to the autoimmune disorder data, additional auxiliary GWAS summary results were obtained from the Early Growth Genetics (EGG) Consortium; see Table~\ref{tab:egg}. The EGG Consortium conducts large GWAS meta-analyses of traits related to childhood growth in humans. It is reasonable to suspect that some of the loci that cause pediatric autoimmune disorders may also affect the early growth traits listed in Table~\ref{tab:egg}. Integrative classifiers were applied to leverage the large sample sizes in the EGG studies to potentially improve the accuracy of predicting autoimmune disease risks.

Subjects were genotyped using Illumina HumanHap550 and Human610 BeadChip arrays, and only variants on autosomal chromosomes typed on both platforms were considered in this analysis. Similar to \citet{shi2016novel}, SNPs were pruned using the \verb|--indep-pairwise| tool in the software PLINK \citep{purcell2007plink} such that no SNPs within a 50 base pair window had $r^2>0.01$. Pruning was performed using reference CEU genotype data from release 23 of the HapMap Project \citep{gibbs2003international} and left 9,491 SNPs. In each of the datasets in Table~\ref{tab:paid}, missing genotypes were then imputed by estimating the minor allele frequency of the corresponding SNP and then using the estimate to randomly generate genotypes assuming Hardy-Weinberg equilibrium. More sophisticated imputation strategies can also be employed. Before applying the classifiers, SNPs with minor allele frequencies below 1\% and/or found not to be in Hardy-Weinberg equilibrium at a $p$-value threshold of $10^{-3}$ were dropped from analysis.

Each classifier described in Section~\ref{sec:compared} was applied to all 130 possible target-auxiliary disease pairs. Gender was always included as a predictor. It was incorporated into PRS, adaptive PRS, and NEBULA using the method described in~\eqref{eq:nebula} in Section~\ref{sec:neb}, and it was not penalized in adaptive PRS or adaptive lasso. The Koenker-Mizera estimator~\eqref{eq:km} used in NEBULA was implemented using 40 equally-spaced points to construct the grids $\Pi_y,y\in\{0,1\}$ and $\Lambda$.

\begin{figure}
  \centerline{\includegraphics[width=\textwidth]{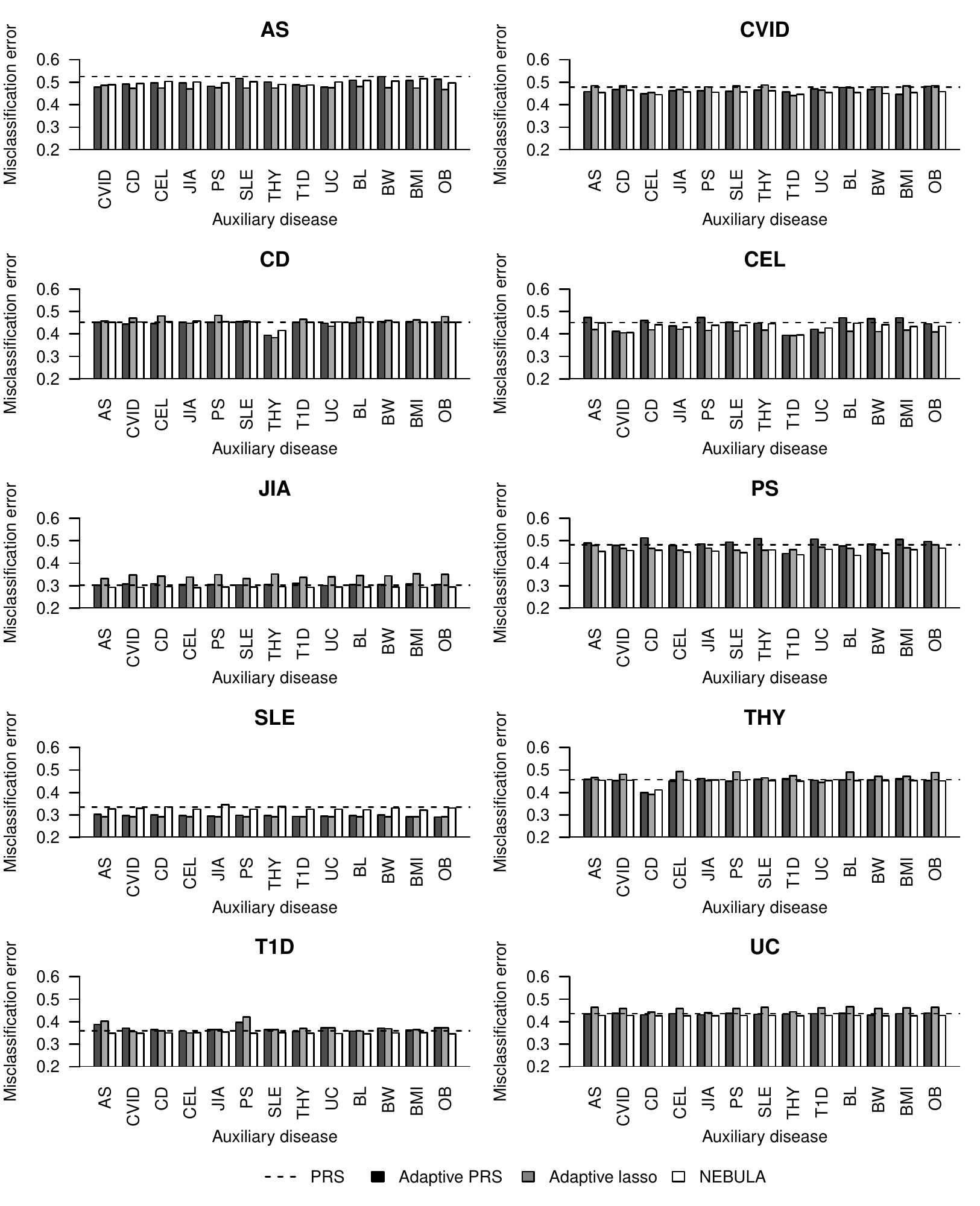}}
  \caption{\label{fig:paid}Average misclassification rates for target pediatric autoimmune diseases, using other autoimmune disease and EGG consortium GWAS results as auxiliary data. See Tables~\ref{tab:paid} and \ref{tab:egg} for abbreviation definitions.}
\end{figure}

The average misclassification rates over the 50 replications are plotted in Figure~\ref{fig:paid}. They show that integrative classification can indeed be effective in improving prediction performance relative to the baseline PRS classifier. The most striking examples occur with Crohn's disease (CD) as the target and thyroiditis (THY) as the auxiliary, and with THY as the target and CD the auxiliary. For these pairs all three integrative methods have substantially lower misclassification rates than non-integrative PRS.

Figure~\ref{fig:paid} also shows that the proposed NEBULA classifier can outperform the other two integrative procedures. For example, when predicting common variable immunodeficiency disorder (CVID) and psoriasis (PS), NEBULA had the lowest misclassification errors among all classifiers for all auxiliary diseases. Of particular interest are the results of using the EGG Consortium's childhood birth length (BL) GWAS summary statistics as auxiliary data to predict PS. Here, NEBULA outperformed adaptive PRS, and adaptive lasso and gave a nearly 10\% improvement in prediction performance relative to non-integrative PRS.

When predicting systematic lupus erythematosus (SLE), NEBULA was outperformed by the other integrative classifiers. It turns out that SLE can be very accurately predicted using gender alone as a single predictor. This corresponds to the sparse predictor setting, and as discussed in the simulations in Section~\ref{sec:sims}, it is not unexpected that adaptive PRS and adaptive lasso can outperform NEBULA when important predictors are very sparse.

On the other hand, as mentioned in Sections~\ref{sec:compared} and \ref{sec:sims}, a disadvantage of adaptive PRS and adaptive lasso is that they can sometimes perform much worse than the baseline non-integrative PRS when the target and auxiliary diseases are highly genetically heterogeneous. This phenomenon was observed when predicting many of the disorders, such as THY, juvenile idiopathic arthritis (JIA), PS, and T1D. In contrast, NEBULA was never much worse than baseline in any of the analyses.

\section{\label{sec:discussion}Discussion}
This paper has so far discussed the integration of auxiliary GWAS summary statistics $T_j$ to improve genetic risk prediction, as the $T_j$ provide latent annotation information for each SNP. However, in some cases important annotations are directly observed, for example if the SNP lies in a DNase I hypersensitive site. It is simple to extend NEBULA~\eqref{eq:nebula} to accommodate directly observed annotations. Let $I_j\in\{0,1\}$ annotate the $j$th SNP. Then the prior assumption~\eqref{eq:G} on minor allele frequencies $\pi_{yj},y\in\{0,1\}$ becomes $(\pi_{0j},\pi_{1j})\mid I_j=0\sim G_0$ and $(\pi_{0j},\pi_{1j})\mid I_j=1\sim G_1$. The $G_y,y\in\{0,1\}$ can again be estimated nonparametrically, and a classifier similar to NEBULA can be derived. The advantage of this formulation is that it does not require perfect concordance between the SNPs predictive of the target disease and those with $I_j=1$. Furthermore, it is also simple modify NEBULA to simultaneously integrate both $T_j$ and $I_j$. These extended classifiers are implemented in the R package \verb|ssa|.

In principle the NEBULA framework can be applied to multiple sources of auxiliary summary statistics $T_j^k$, $k=1,\ldots,K$. This would necessitate a multivariate prior distribution on $(\pi_{0j},\pi_{1j},T_j^1,\ldots,T_j^K)$, from which the corresponding Bayesian classifier can be derived. However, nonparametric estimation of this multivariate prior will be both theoretically and computationally challenging. A possible solution is to develop convenient closed-form estimators, which may not be optimally adaptive to the true prior $G$ but may still have good performance. This is an interesting direction of future research.

Even in the present case, nonparametric estimation of a trivariate prior distribution is computationally challenging. Faster strategies would make feasible genetic risk prediction using genome-wide SNP data without LD pruning. One approach would be to limit the support of the estimated prior. For example, the Koenker-Mizera estimator described in Section~\ref{sec:imp} supports the $(\pi_{0j},\pi_{1j})$ on the grid $\Pi_0\times\Pi_1$. However, most SNPs frequencies are likely to be similar between cases and controls, such that $\pi_{0j}\approx\pi_{1j}$ for most $j$. Thus most grid points in $\Pi_0\times\Pi_1$ should have zero mass. Enforcing this can reduce the number of grid points and speed up estimation of the prior $G$. Further work in this direction is necessary.

\section*{Acknowledgments}
The author thanks Drs. Hakon Hakonarson, Brendan J. Keating, Yun Li, and Julie Kobie for providing the pediatric autoimmune disease data, and Dr. Lee Dicker for helpful discussion. Data on the birth length, birth weight, childhood obesity, and childhood BMI traits have been contributed by EGG Consortium and have been downloaded from \verb|www.egg-consortium.org|. The research of Dave Zhao was supported in part by NSF grant DMS-1613005 and by a grant from the Simons Foundation (\#SFLife 291812).

\bibliographystyle{abbrvnat}
\bibliography{refs}

\end{document}